\documentclass[twocolumn]{aastex63}

\usepackage{lineno}
\usepackage{CJK}
\usepackage{amsbsy}
\usepackage{multirow}
\usepackage{bm}
\usepackage{ulem}

\received{mm dd, yyyy}
\revised{mm dd, yyyy}
\accepted{mm dd, yyyy}
\submitjournal{ApJ}
\shorttitle{Quenching of massive disk galaxies}
\shortauthors{Xu et al.}
\graphicspath{{./}{figure_QMSG_PDF/}}

\begin{document}
\begin{CJK*}{UTF8}{gbsn}

\title{Quenching of Massive Disk Galaxies in the IllustrisTNG Simulation
\footnote{Released on mm, dd, yyyy}}

\correspondingauthor{Xi Kang}
\email{xuyingzhong@zju.edu.cn,
       kangxi@zju.edu.cn}

\author[0000-0003-4823-1898]{Yingzhong Xu(徐迎中)}
\affiliation{Zhejiang University-Purple Mountain Observatory Joint Research Center for Astronomy, Zhejiang University, Hangzhou 310027, China}

\author[0000-0003-2341-9755]{Yu Luo(罗煜)}
\affiliation{Purple Mountain Observatory, 10 Yuan Hua Road, Nanjing 210034, China}
\affiliation{School of Astronomy and Space Sciences, University of Science and Technology of China, Hefei 230026, China}
\affiliation{National Basic Science Data Center, Zhongguancun South 4th Street, Beijing 100190, China}

\author[0000-0002-5458-4254]{Xi Kang(康熙)}
\affiliation{Zhejiang University-Purple Mountain Observatory Joint Research Center for Astronomy, Zhejiang University, Hangzhou 310027, China}
\affiliation{Purple Mountain Observatory, 10 Yuan Hua Road, Nanjing 210034, China}

\author{Zhiyuan Li(李志远)}
\affiliation{School of Astronomy and Space Science, Nanjing University, Nanjing 210093, China}
\affiliation{Key Laboratory of Modern Astronomy and Astrophysics, Nanjing University, Nanjing 210023, China}

\author{Zongnan Li(李宗男)}
\affiliation{School of Astronomy and Space Science, Nanjing University, Nanjing 210093, China}
\affiliation{Key Laboratory of Modern Astronomy and Astrophysics, Nanjing University, Nanjing 210023, China}

\author{Peng Wang(王鹏)}
\affiliation{Leibniz-Institut f\"ur Astrophysik Potsdam (AIP), An der Sternwarte 16, D-14482 Potsdam, Germany}

\author{Noam Libeskind}
\affiliation{Leibniz-Institut f\"ur Astrophysik Potsdam (AIP), An der Sternwarte 16, D-14482 Potsdam, Germany}
\affiliation{University of Lyon; UCB Lyon 1/CNRS/IN2P3; IPN Lyon (IPNL), France}

\begin{abstract}
A rare population of massive disk galaxies have been found to invade the red sequence dominated by early-type galaxies. These red/quenched massive disk galaxies  have recently gained great interest into their formation and origins. The usually proposed quenching mechanisms, such as bar quenching and environment quenching, seem not suitable for those bulge-less quenched disks in low-density environment. In this paper, we use the TNG300 simulation to investigate the formation of massive quenched central disk galaxies. It is found that these galaxies contain less gas and harbor giant supermassive black holes(SMBHs) (above $ 10^{8}M_{\odot}$) than their star forming counterparts. By tracing their formation history, we found that quenched disk galaxies formed early and preserved disk morphology for cosmological time scales. They have experienced less than one major merger on average and it is mainly mini-mergers  (mass ratio $<$1/10) that contribute to the growth of their SMBHs. In the IllustrisTNG simulation the black hole feedback mode switches from thermal to kinetic feedback when the black hole mass is more massive than $\sim 10^{8}M_{\odot}$, which is more efficient to eject gas outside of the galaxy and to suppress further cooling of hot gaseous halo. We conclude that kinetic AGN feedback in massive red/quenched disk galaxy is the dominant quenching mechanism.

\end{abstract}

\section{Introduction} \label{sec:intro}
Our knowledge of galaxy formation has greatly benefited from large sky surveys e.g. SDSS \citep{2000AJ....120.1579Y}, which have provided large sample of galaxies across wide range of mass, morphology and environment. These rich data allows us to infer the statistical properties of galaxy population \citep[e.g.][]{2003MNRAS.341...33K}, among which the most distinct feature is the bi-modal distribution of galaxy population in the color-magnitude or morphology-mass diagram  \citep[e.g.][]{2003ApJ...594..186B,2014MNRAS.440..889S}. This bi-modality states that there are two main branches of galaxy population, one is the red sequence consist of passive, bulge-dominated massive galaxies, while the blue sequence consists of star forming, disk-dominated low-mass galaxies. \par

Reproducing the two distinct sequences of galaxy population from theoretical modelling is not easy. In particular, the quenching of massive galaxies has long been the main motivation of semi-analytical model \citep[e.g.][]{2003ApJ...599...38B,2006MNRAS.365...11C,2006ApJ...648..820K}  hydro-dynamical simulations  \citep[e.g.][]{2014MNRAS.444.1518V}. Though it is yet unclear in the physics of quenching in detail, it is commonly believed that quenching of massive galaxy is related to the central SMBHs. Frequent mergers, especially major mergers, lead to formation of galaxy bulge \citep[][]{1972ApJ...178..623T} and strong torque on gas lead to rapid feeding of black hole at the center. The strong feedback from high-accretion state of black hole (QSO phase) can effectively sweep up the cold gas from the galaxy \citep[e.g.][]{2005Natur.433..604D}, and subsequent feedback from radio AGN (in low-accretion state) can heat the hot gas in the dark matter halo, suppressing the further cooling of fresh gas for a cosmological time-scale \citep[e.g.][]{2005MNRAS.361..776S}. Such a merger-driven formation scenario for growth of both galaxy bulge and SMBH fits well with the observed galaxy bulge-black hole mass relation \citep[e.g.][]{2001AIPC..586..363K}, and it has been implemented , though differing significantly in detail, in semi-analytical model \citep[e.g.][]{2000MNRAS.311..576K,2006MNRAS.365...11C,2006ApJ...648..820K,2006MNRAS.370..645B,2008MNRAS.391..481S,2009Natur.460..213C} and the state-of-the-art hydro-dynamical simulation of galaxy formation \citep[e.g.][]{2007MNRAS.380..877S,2008ApJS..175..356H,2014Natur.509..177V,2015MNRAS.446..521S,2018MNRAS.473.4077P} to well reproduce the red sequence of galaxies.  However, more recent evidence from both  observations\citep[e.g.][]{2008ApJ...687..216G,2010ApJ...721...26G,2011ApJ...742...68J,2011ApJ...734..121S,2011ApJ...741L..11C,2011ApJ...727L..31S,2012MNRAS.425L..61S,2012ApJ...744..148K,2013MNRAS.429.2199S,2014ApJ...782...22B,2014MNRAS.440.2944K,2017MNRAS.470.1559S,2019MNRAS.489.4016S,2020ApJ...899...82B,2021MNRAS.507.3985S} and simulations\citep[e.g.][]{2018MNRAS.476.2801M,2020MNRAS.494.5713M} have found that SMBHs and AGN activities are also seen in disk-dominated galaxies, suggesting that secular processes other than major mergers, such as minor mergers and smooth gas accretion, may also play roles in driving the growth of both SMBHs and galaxies. \par

In addition to the dominant population of early-type galaxies at the massive end ($>10^{10}M_{\odot}$) of the red sequence, a rare population of disk galaxies also intrude the red sequence at the massive end. The red disk galaxy population has attracted great attention recently \citep[e.g.][]{2019ApJ...884L..52Z,2021ApJ...911...57Z,2020arXiv201113749Z,2021SCPMA..6479811X}. In fact red disk galaxies have been known for several decades \citep[e.g.][]{1976ApJ...206..883V,1998ApJ...497..188C}. It was found that most red disk galaxies are really passive that their star formation has quenched \citep[e.g.][]{2013MNRAS.432..359T,2020ApJ...897..162G}. Most red disks preferentially reside in intermediate/high density regions and they  also show distinct sign of  bars \citep[e.g.][]{2010MNRAS.405..783M,2019ApJ...883L..36H,2020MNRAS.491..398M,2021MNRAS.501.3309Z}. \par

The formation and quenching mechanism of red disk galaxy is still not clear. It is widely speculated that both internal structure and external environment could lead to the quenching of red disk\citep[e.g.][]{2018MNRAS.474.1909F} . Red disk with significant bars are quenched by internal process, such as bar quenching, while those in high/intermediate environment density may be quenched due to the environment effect, such as ram-pressure stripping or strangulation. However, such quenching mechanisms may not apply to recent findings of massive disk galaxies lacking clear signature of  bar and in low-density environment \citep{2020MNRAS.496L.116L,2021SCPMA..6479811X}.  
For these galaxies, some additional quenching mechanisms have to be invoked, such as suppression of cooling from hot gaseous halo\citep[e.g.][]{2018NatAs...2..695M}.  AGN feedback is a very natural choice. Strong AGN feedback, under the merger-driven paradigm, seems not applicable  in massive bulgeless galaxies as they are not expected to possess giant SMBHs. But some recent observations do confirm the presence of strong AGN in massive disk dominant galaxies\citep[e.g.][]{2010ApJ...721...26G,2011ApJ...741L..11C,2014ApJ...782...22B}  and even in the massive bulgeless ones\citep[e.g.][]{2008ApJ...687..216G,2011ApJ...727L..31S,2012ApJ...744..148K,2012ApJ...761...75S,2013MNRAS.429.2199S,2017MNRAS.470.1559S,2020ApJ...899...82B}. Interestingly, it was the findings of those merger-free objects that challenged the merger-driven mechanism\citep[e.g.][]{2010ApJ...721...26G,2011ApJ...741L..11C,2011ApJ...742...68J,2013MNRAS.429.2199S}.\par

Recently, \cite{2020MNRAS.496L.116L} study the red disk galaxy using the model galaxies from the IllustrisTNG simulation \citep[e.g.][]{10.1093/mnras/stx3112,2018MNRAS.480.5113M}.  They found eight quenched disk galaxies and have shown that these galaxies have massive black hole mass ($>10^{8}M_{\odot}$), about $\sim 1\ \mathrm{dex}$ higher than the predictions from the classical bulge mass-black hole mass relation. \cite{2020MNRAS.496L.116L} speculated that it was the AGN feedback that shut off the cooling of hot halo gas, thus the disk galaxies are kept  quenched. However, since their selection of quenched disks is limited to galaxies with stellar masses larger than $10^{11}M_{\odot}$, the sample is too small (only eight galaxies) for statistical analysis.\par

 In this work, we use the IllustrisTNG simulation and select quenched disk galaxies with stellar mass larger than $10^{10.5}M_{\odot}$. Compared with the work of \cite{2020MNRAS.496L.116L}, we have a much larger samples and it enables us to study the properties and formation history of quenched disk galaxies in more detail, and in particular, we focus on the quenching mechanism of these red disk galaxies. This paper is organised as follows: 
in Section \ref{sec:sample}, we introduce selection of galaxy sample from the simulation and present their properties in Section \ref{sec:analyse-samples}.
We explore the evolution of gas, SMBHs and specific star formation rates in Section \ref{sec:Causal relationship}.  We go further to analyse the growing of SMBHs in Section \ref{sec:grow-bh}. In Section \ref{sec:conclusion}, we give our conclusions and present some discussion.

\section{SAMPLE SELECTION} \label{sec:sample}

In this section we shortly highlight a few implementation of the IllustrisTNG simulation  which we believe are relevant to our study and our selection of model galaxy. We select two samples at $z=0$. One is the massive quenched central galaxies(named SampleQ), the other is the massive star-forming central disk galaxies(named SampleF). Their comparison gives us a clue as to why some central disk galaxies are quenched while others are not.
\subsection{IllustrisTNG}\label{sec:tng}
The IllustrisTNG(hereafter TNG)\footnote{\url{https://www.tng-project.org/}} project is a series of cosmological magnetohydrodynamical simulations which have produced many wonderful results \citep[e.g.][]{10.1093/mnras/stx3112,2018MNRAS.480.5113M} which improved our
understanding about galaxy formation.
There are 3 simulation boxes with 3 different resolutions for TNG simulations. In this work, we selected samples from the simulation TNG300-1\citep[][]{2019ComAC...6....2N,2018MNRAS.475..648P,2018MNRAS.475..676S,2018MNRAS.475..624N,2018MNRAS.477.1206N,2018MNRAS.480.5113M} with the largest simulation box (roughly $(300\ \mathrm{Mpc})^{3}$) and high resolution,  as it could provide a large sample of massive galaxies ($M_{\star} \geq 10^{10.5} M_{\odot}$).
It adopted the \citep{2016A&A...594A..13P} best fit cosmological parameters: dark energy 
density $\Omega_{\Lambda} = 0.6911$, matter density $\Omega_{m} = 0.3089$, baryon density $\Omega_{b} = 0.0486$, Hubble constant $H_{0} = 67.74kms^{-1}Mpc^{-1}$, normalisation $\sigma_{8} = 0.8159$ and the
spectral index $n_{s} = 0.9667$.

In the following we list some physical prescriptions of the TNG simulations related to our investigation: 

\begin{itemize}
\item Merger History.
 In the TNG simulations, galaxies live in the halos and subhalos which are identified by using the Friends-of-Friends(FOF) \citep{2001MNRAS.328..726S} and SUBFIND algorithms \citep{2005Natur.435..629S}. The SubLink \citep{2015MNRAS.449...49R} merger tree record the formation history of the galaxies.   The SubLink is a algorithm used to link subhalos at different snapshots and to construct the merger trees. For a subhalo A with $N_{A}$ member particles at snapshot $S_{A}$, it has a main progenitor and a sole descendant at $S_{A}-1$ and $S_{A}+1$ respectively. In general, the member particles of subhalo A at $S_{A}$ are contained in different subhalos at $S_{A}-1$ which are all defined to be progenitors of subhalo A. The main progenitor is defined as the one which contains the maximum number of particles belong to subhalo A. Similarly, the descendant is defined as the halo at $S_{A}+1$ which contains the maximum number of member particles belong to subhalo A. The main progenitor and descendant is then used to construct the merger tree. For more detail, we refer the readers to the paper by \cite{2015MNRAS.449...49R}.

\item Hydrogen Gas. The gas cell is able to form stars when its hydrogen number density is above a threshold($\sim 0.1 cm^{-3}$) according to the star formation model \citep{2003MNRAS.339..289S} which is used in the TNG simulation. The hydrogen number density is $n_{H} = \frac{X_{H}\rho}{m_{p}}$.
where $X_{H}$ is the abundance of hydrogen, $\rho$ is the mass density of the cell, $m_{p}$ is the mass of proton. The star-forming gas($n_{H}>0.1 cm^{-3}$) which is dominated($\gtrsim 85\%$, refer to the first figure in \cite{2003MNRAS.339..289S}) by cold phase($T=10^{3}K$) can be treated as cold gas. It follows the effective equations of state. For the non-star-forming gas($n_{H}<0.1 cm^{-3}$), it follows the ideal-gas equations of state and the temperature is at most cooled to $10^{4}$ K(cool or hot gas).

\item SuperMassive Black Hole. When the mass of a FOF halo, without a SMBH, is above $\sim 7\times 10^{10}M_{\odot}$, a SMBH seed($\sim 10^{6}M_{\odot}$) is put into the center  (where the potential is minimal) of the halo. There are two kinds of feedback mode for the SMBH: thermal and kinetic mode. When the black hole growth is in a high accretion state, 
the feedback energy will be injected into the surrounding gas as thermal energy, while in a low accretion state, momentum rather than thermal energy will be transmitted to the surroundings in a random direction driving a strong wind. 
In TNG simulations, when the mass of the SMBH $M_{bh}$ exceeds $\sim 10^{8}M_{\odot}$ the accretion mode will
switch from thermal to kinetic mode. We recommend readers who are interested in the black hole model used in the TNG to the paper by \cite{2017MNRAS.465.3291W}.
\end{itemize}

\subsection{Selecting model  galaxies}\label{sec:selecting-method}
We investigate the quenched massive disk central galaxies from TNG300-1 data at $z=0$, therefore we use three quantities to select galaxy sample: the stellar mass  $M_{\star}$, the specific star formation rate,  $\mathrm{sSFR}=\frac{\mathrm{SFR}}{M_{\star}}$($\mathrm{SFR}$ represents the instantaneous 
star formation rate, which is a direct output in the TNG simulation) and the ratio of spheroid stellar mass to total 
stellar mass of galaxy, labelled as $\mathrm{S/T}$.\par 

Firstly, we select central galaxy  defined as the most massive member galaxy in a group and calculate the above three quantities within $3\mathrm{R_{e}}$ ($R_{e}$ is the stellar half-mass radius) relative to the centre of the galaxy. 
The calculation of $M_{\star}$ and $\mathrm{sSFR}$ is straightforward.  For $M_{\star}$, we summed the mass of every star particles within $3\mathrm{R_{e}}$. For $\mathrm{sSFR}$
we summed the SFR(the direct output of the simulation) of every gas cells within $3\mathrm{R_{e}}$ and divided it by the stellar mass $M_{\star}$. 
 
\begin{figure}[ht!]
    \plotone{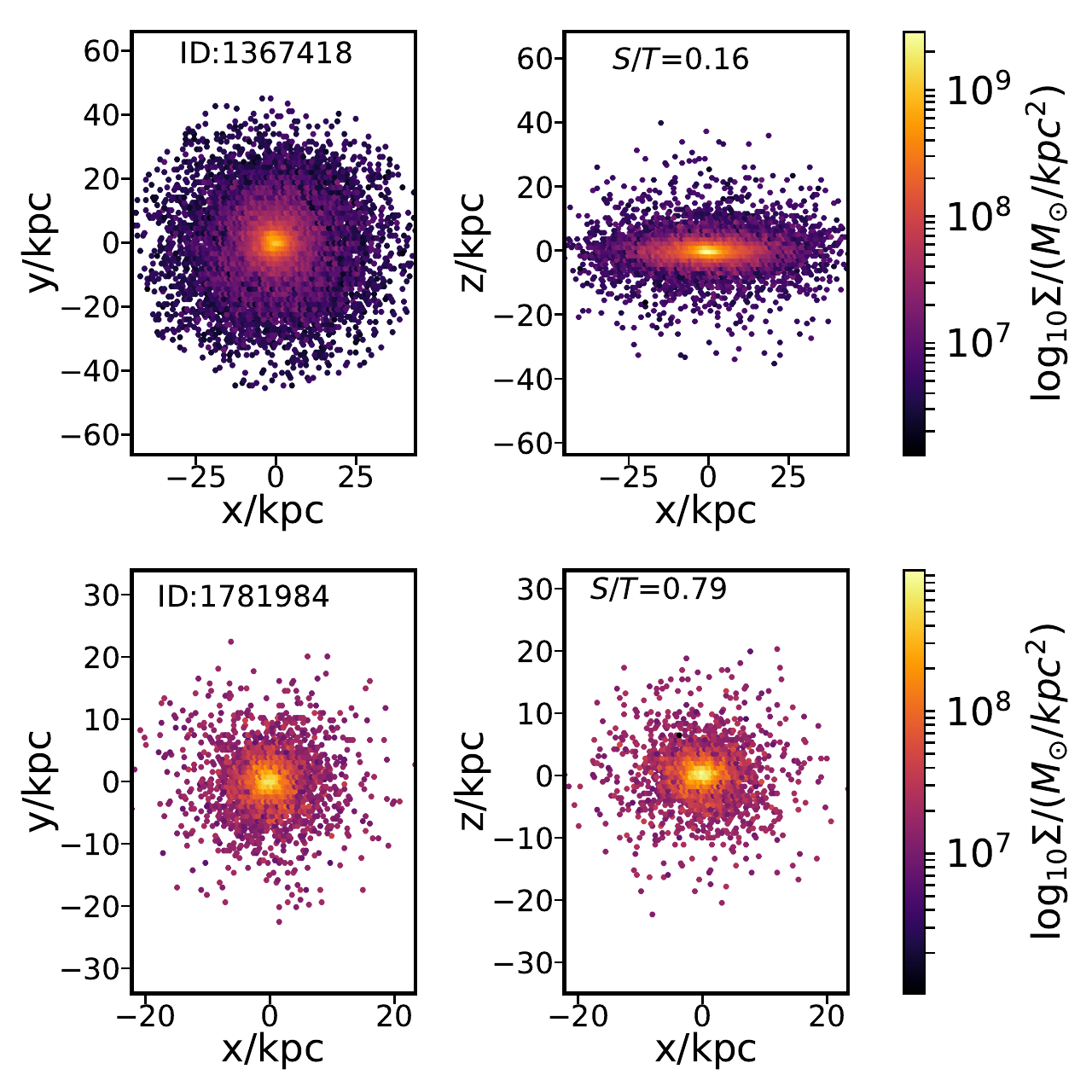}
    \caption{The top and bottom row show the face-on(left) and edge-on(right) stellar surface mass density distribution of two randomly selected galaxies, respectively. Their TNG ID and the spheroid to total mass ratio(S/T), calculated using the methods we introduced, were also labeled in the  panels. In each image we only considered star particles within $6\mathrm{R_{e}}$. Obviously, the KD methods we used do pick out the disk galaxies.}
    \label{fig:eg_ratio}
\end{figure}

While for $\mathrm{S/T}$, we need to decompose the stellar mass into the disk and spheroid components.
The classical decomposition method is the photometric decomposition which relies on 
the empirical assumption that disk and bulge components of a galaxy have different surface brightness profiles.
However, there are some more suitable methods than the photometric decomposition for simulation data, such as the kinematic decomposition(KD) which decomposes a galaxy according to the distribution of a characteristic quantity $\epsilon$. There are a few definitions of $\epsilon$ \citep[e.g.][]{2003ApJ...597...21A,2003ApJ...591..499A,2009MNRAS.396..696S,2012MNRAS.421.2510D,10.1093/mnras/stx3112}. In this work, we choose the same definition of $\epsilon$ as used in \cite{2019MNRAS.487.5416T}.   We note that there is no particular reason we selected the method of \cite{2019MNRAS.487.5416T} over other methods, only because this decomposition is more faster as it requires no calculation of potential energy of each particle as required by other kinematic methods, and as shown by \cite{2019MNRAS.487.5416T} the obtained S/T is well correlated with that from other kinematic methods.
The definition we used is $\epsilon=\frac{j_{z}}{j}$, where $j$ is the angular momentum of a stellar particle and $j_{z}$ is the component of angular momentum projected to the direction of total stellar angular momentum of a galaxy within $3\mathrm{R_{e}}$. So the mass of the spheroid component, $M_{s}$, is twice the sum of the mass of stellar particles with $\epsilon \leq 0$, under simple assumption that the bulge stars move randomly with a symmetrical distribution. Then we got $S/T=\frac{M_{s}}{M_{total}}$, here $M_{total}$ is the the total mass of all the stellar particles within $3\mathrm{R_{e}}$.   
It is noted that $S/T$ could be larger than $1$ as we simply double the stellar mass with $\epsilon \leq 0$, hence we removed those galaxies with $S/T >1$ which are not realistic (only $\sim 1\%$ of the total $22417$ massive central galaxies in our sample). In Figure \ref{fig:eg_ratio} we randomly chose two galaxies(showed in the first and second row respectively) and plotted their face-on(left) and edge-on(right) stellar surface mass density distribution. The spheroid to total ratio(S/T) which were calculated by using the above KD methods were also labeled in the panels. This figure show that the methods we used in this paper do work well.

For our study in this work, we define a galaxy to be bulgeless if $\mathrm{S/T}<0.25$, and a galaxy with $\mathrm{sSFR}<10^{-11} yr^{-1}$ to be quenched. For our purpose we finally select two samples of galaxies, one is quenched disk galaxies with $M_{\star}>10^{10.5}M_{\odot}, \mathrm{sSFR}< 10^{-11} yr^{-1}\mathrm{and}\ \mathrm{S/T}<0.25$, named as SampleQ, and the other is star forming disk galaxies with $M_{\star}>10^{10.5}M_{\odot}, \mathrm{sSFR}> 10^{-11} yr^{-1}\mathrm{and}\ \mathrm{S/T}<0.25$, named as SampleF. From the TNG simulation, we get 545 and 658 members, respectively for the two samples. Considering the resolution limit of the simulation, the star formation rate (SFR) will be set to 0 when the SFR is unresolved. Thus, here we set $\mathrm{sSFR} = 10^{-20} yr^{-1}$ when $\mathrm{SFR}=0\ M_{\odot}yr^{-1}$.


\begin{figure}[ht!]
    \plotone{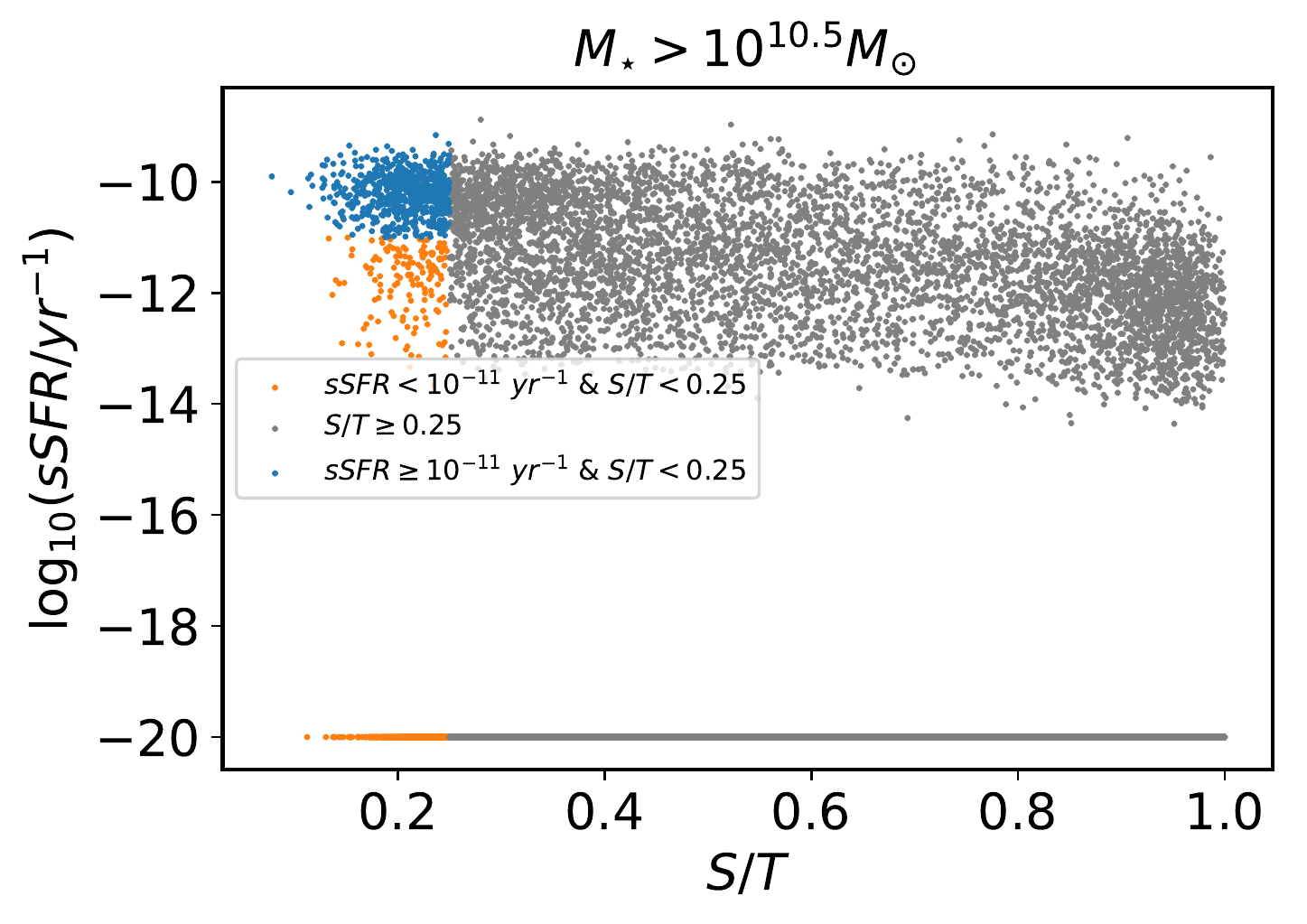}
    \caption{The S/T and sSFR distribution for massive($M_{\star}>10^{10.5}M_{\odot}$) central galaxies. blue points 
    belong to SampleF, orange points belong to SampleQ, gray points are the rest. For those galaxies with $\mathrm{SFR}=0\ M_{\odot}yr^{-1}$, we set $\mathrm{sSFR} = 10^{-20}yr^{-1}$.}
    \label{fig:ssfr_s/t}
\end{figure}

In Figure \ref{fig:ssfr_s/t} we show the distribution of TNG model galaxies on the specific star formation and morphology diagram. Here all central galaxies with $M_{\ast}>10^{10.5}M_{\odot}$ are shown, while the blue points are for SampleF and orange points are for SampleQ. In general, the simulation is consistent with the observational trend that at given mass, bulge-dominated galaxies have lower star formation rate and disk-dominated galaxies have higher star formation rate \citep[e.g.][]{2003MNRAS.341...54K,2015MNRAS.454.1886S}. In the following sections we mainly focus on disk-dominated galaxies (colored dots) and investigate why some of them are quenched and others are not.

\section{The gas content and black hole of quenched massive disk galaxies}\label{sec:analyse-samples}
In this section, we examine the differences between the SampleQ and SampleF in some physical properties, such as the gas content and the central black hole mass, as they are thought to be the main factors leading to the quenching of disk galaxies \citep[e.g.][]{2020MNRAS.496L.116L}. Investigation of other properties, such as galaxy structure between SampleQ and SampleF, requires simulation with more high-resolutions which is not available at the moment.

\subsection{The gas content}
\begin{figure}
    \plotone{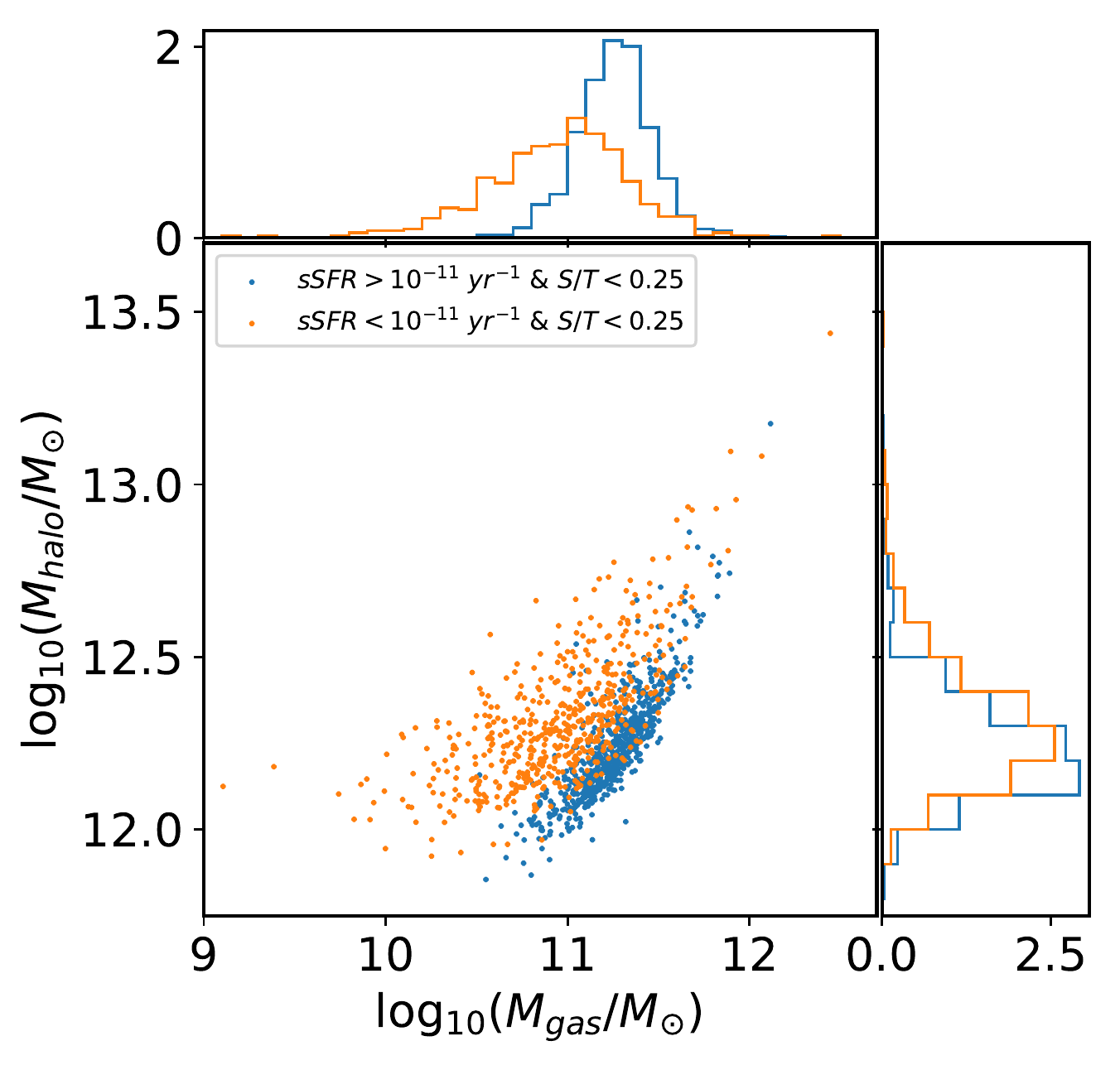}
    \caption{This scatter plot shows the relation between the halo virial mass $M_{halo}$ and the gas mass $M_{gas}$ and the two marginal 
            distributions are also displayed on the top and right. The orange points(i.e.quenched galaxies) own fewer gas than the star-forming one with similar halo virial mass.}
    \label{fig:halo}
\end{figure}

Both hot and cold gas are related to star formation activity in a galaxy, as the former is the source of cooling and the latter is direct fuel for star formation. In Figure \ref{fig:halo},  we show the halo gas  mass and halo virial mass distribution of those two samples, where the total gas is referred to all the gas cell within halo, including both cold and hot, and the halo virial mass $M_{halo}$ is the total mass of a sphere inside which the mean density is 200 times the critical density. It's clearly seen that, at the fixed halo virial mass, the quenched sample (orange dots) have lower halo gas  mass than the star forming sample (blue dots), while two samples have similar halo virial mass distribution (right small panel). The same phenomenon was also found
by \cite{2020MNRAS.491.4462D}.

\begin{figure}
    \plotone{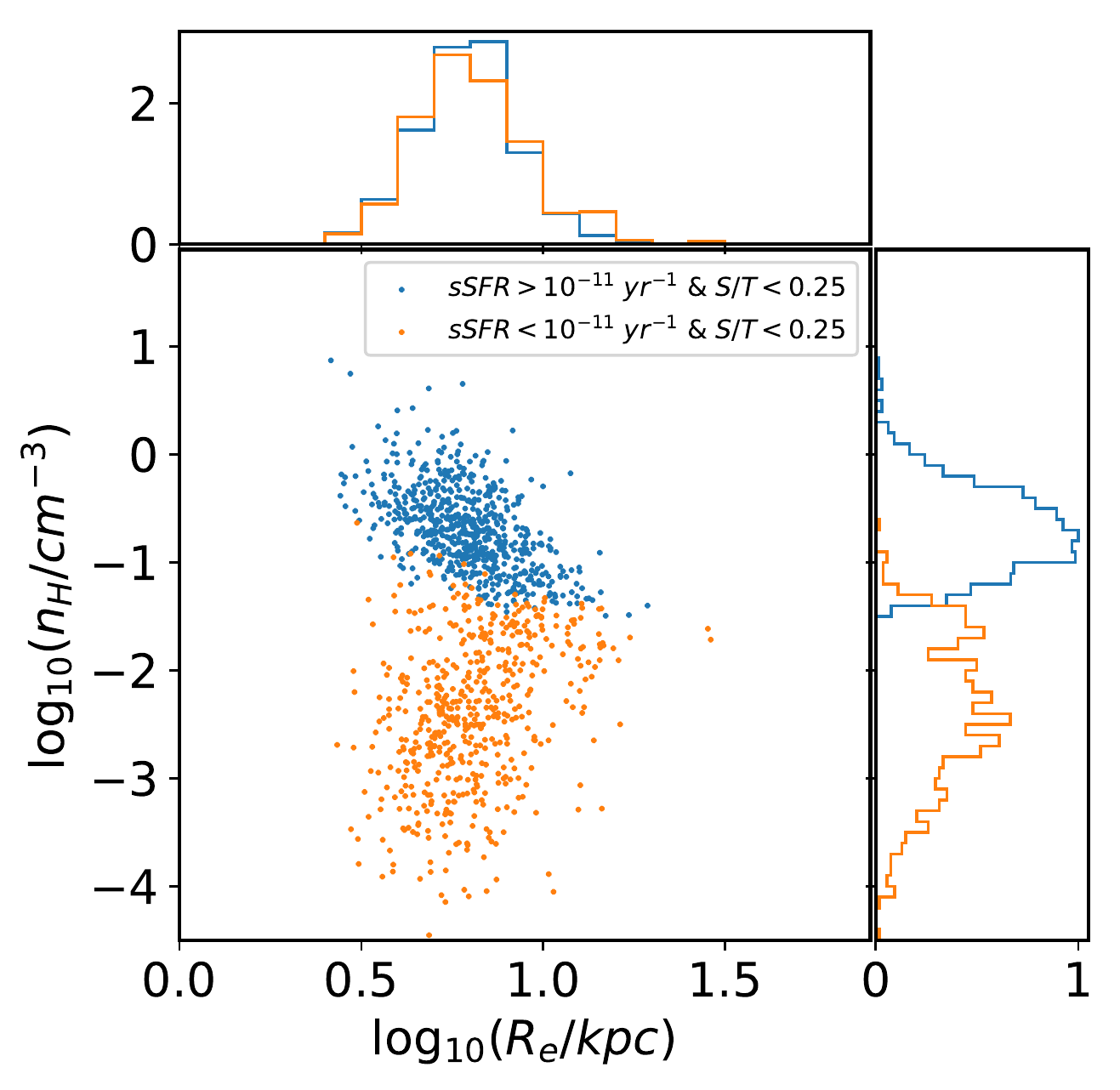}
    \caption{This figure is similar to the Figure \ref{fig:halo} except about the relation between the mean gas density 
            $n_{H}$ and the stellar half-mass radius $\mathrm{R_{e}}$. Obviously, the SampleQ lack star forming gas($n_{H}>0.1cm^{-3}$) while their size($R_{e}$) are similar to that of the SampleF.}
    \label{fig:numberDensity_Re}
\end{figure}

As cold gas, mainly in hydrogen, is the direct gas reservoir for star formation, in Figure \ref{fig:numberDensity_Re} we show the distribution of the mass-weighted mean hydrogen gas density $n_{H}$ within  $3\mathrm{R_{e}}$ to illustrate the cold (hydrogen) gas content. As expected, the mean hydrogen gas densities of SampleQ are systematic lower than $0.1cm^{-3}$, while that of SampleF are higher. In the TNG simulation a gas cell is allowed to form stars only if its density, $n_{H}$, is above the threshold $0.1cm^{-3}$. This means the SampleQ lack hydrogen gas to form stars within $3\mathrm{R_{e}}$. It is also found that the lower hydrogen gas in SampleQ is not due to their higher disk size, as seen from the upper panel the distributions of disk size in SampleQ and SampleF are very similar.        

The above results show that both the total and cold gas are less in SampleQ than in SampleF. As cold gas is from the cooling of hot gas and the gas content of our selected sample galaxies are dominated by hot gas, the deficiency of cold gas in SampleQ is mainly due to their shortage of total gas compared to SampleF. In the following text we do not distinguish the cold gas from the hot gas, but focus on the total gas evolution and investigate which physical process produce less total gas in SampleQ.\par

\subsection{The black hole mass}
As mentioned before, black hole is a key component of modelling in nearly all hydro-dynamical simulations.  Traditionally the mass of black hole in elliptical galaxies is measured using spectroscopic methods, and it was recently found that disk galaxies follow similar black hole mass-bulge (or total) mass relation but with different slope and larger scatter \citep[e.g.][]{2018ApJ...869..113D,2019ApJ...873...85D}. It is not the focus of this work to compare the predicted black hole mass from TNG simulation with the data in detail, but to gain insight on the possible effects of black hole feedback on disk quenching. For more comparison of predicted black hole mass with the data from TNG simulation, the readers are referred to a few recent work\citep[e.g.][]{2020ApJ...895..102L,2020MNRAS.493.1888T}.

\begin{figure}[ht!]
    \plotone{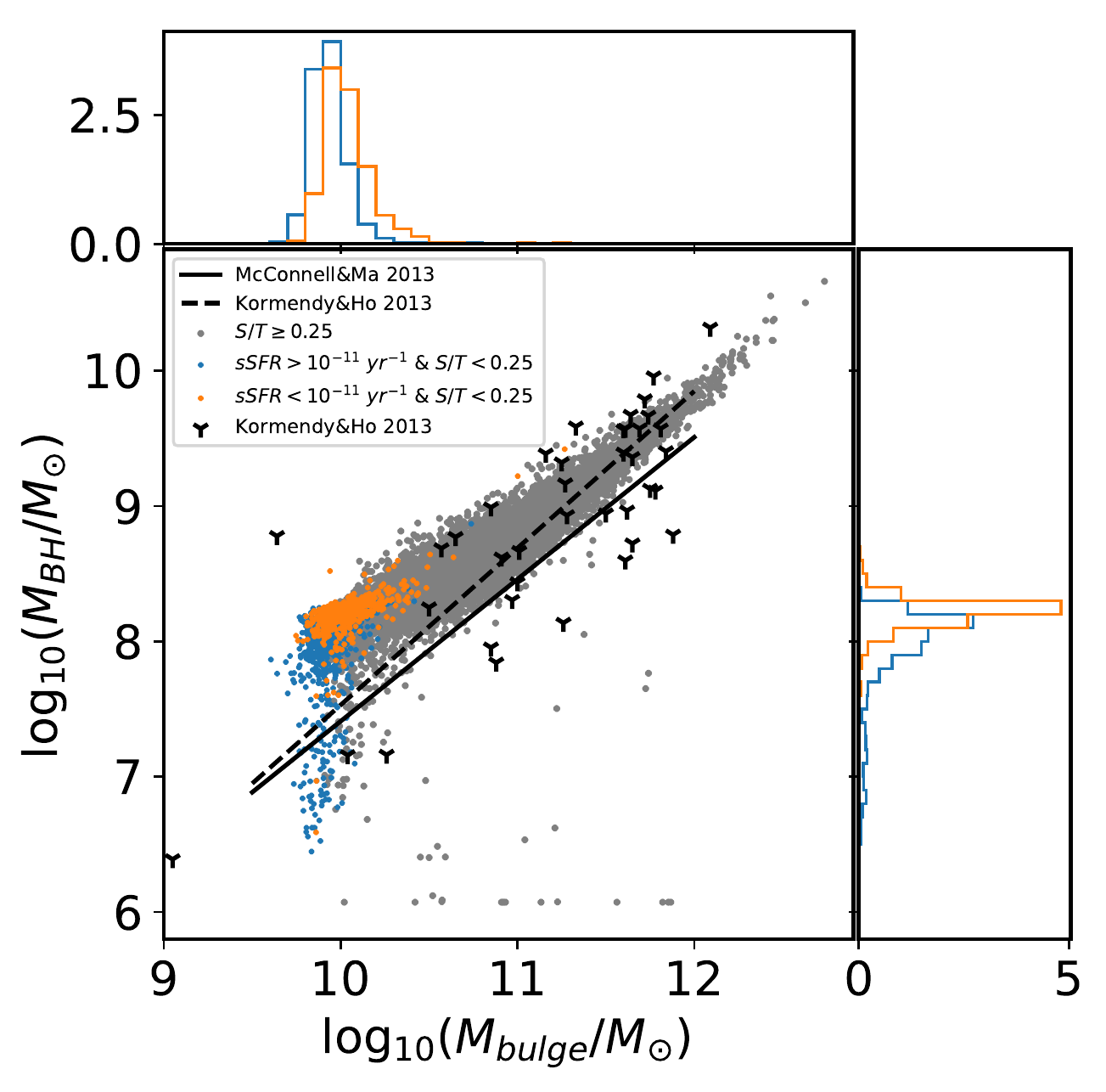}
    \caption{The relation between black hole mass and bulge mass for massive($M_{\star}>10^{10.5}M_{\odot}$) central galaxies. blue points 
    belong to SampleF, orange points belong to SampleQ, gray points are the rest. sampleQ own more massive SMBH($>10^{8}M_{\odot}$) than SampleF with similar bulge mass.
    The black mark and solid line are quoted from \cite{2013ARA&A..51..511K} and the dashed
    line are quoted from \cite{2013ApJ...764..184M}.}
    \label{fig:bh-bulge-full}
\end{figure}

Figure \ref{fig:bh-bulge-full} shows the bulge-SMBH relation.  It is straightforward to see that both the quenched galaxies (orange dots) and star-forming galaxies (blue dots) have similar bulge mass, while the quenched ones have higher black hole mass at given bulge mass. However, it is notable that the simulated galaxies lie systematically above the two classical  bulge-SMBH relations \citep{2013ARA&A..51..511K,2013ApJ...764..184M} (solid and dashed lines).
 We note that such a deviation from the observed bulge-SMBH relation at the low mass end is also seen in the highest resolution TNG simulation with a box size of $(30 h^{-1}\  
\mathrm{Mpc})^{3}$ by \cite{2017MNRAS.465.3291W} who concluded that this may be a result of increasing mass of the seed black hole and the definition of the bulge mass(It is notable that they used the same definition of bulge mass as this work did, except they considered star particles within 0.1 virial radius, $0.1R_{200}$, not the $3\mathrm{R_{e}}$ we used here). But this will not affect our conclusion that galaxies of SampleQ have systematically higher black hole mass than SampleF.

From the above results, we can conclude that the galaxies of SampleQ have larger black hole($>10^{8}M_{\odot}$), less cold and less total gas than those of SampleF, while for other quantities(e.g. the halo virial mass $M_{halo}$, the stellar half-mass radius $\mathrm{R_{e}}$ ), the two samples have similar distributions. \par

\cite{2019MNRAS.485.3783D,2020MNRAS.491.4462D} have found a significant positive correlation between the specific star formation rate(sSFR) and the circumgalactic medium (CGM) mass fraction($f_{CGM}\equiv \frac{M_{gas}}{M_{200}}$, $M_{gas}$ is the total mass of the gas cells within virial radius which are non-star-forming) for central galaxies in EAGLE\citep{2015MNRAS.446..521S,2015MNRAS.450.1937C} and TNG simulations, since the lack of fuel to form star.
Their results also indicated that the low $f_{CGM}$ is due to the SMBH. Particularly for TNG, the AGN 
feedback mode will change from thermal to kinetic when the $M_{\mathrm{bh}}$ is above the threshold($\sim 10^{8}M_{\odot}$) \citep{2017MNRAS.465.3291W}. In the  kinetic mode, the gas can be expelled effectively from the galaxy or even be driven out of the halo\citep[e.g.][]{2020MNRAS.493.1888T,2018MNRAS.479.4056W,2020MNRAS.499..768Z}.

The above quenching mechanism seems applicable to our SampleQ which have both lower gas in the halo and giant SMBH($M_{bh}>10^{8}M_{\odot}$).  However, it is unknown why these quenched disk galaxies have higher black hole mass. It is also interesting to note that there are some galaxies with SMBHs ($\sim 10^{8}M_{\odot}$) but they are still star forming. Thus having SMBH does not necessarily lead to the quenching of a galaxy. The answers may be behind the formation history of the galaxies and their SMBHs. In the following sections we investigate their formation history in detail.


\section{The evolution history of quenched massive disk galaxies} \label{sec:Causal relationship}
We used the SubLink merger tree data of the TNG simulation(see Section \ref{sec:tng}) to investigate the evolution history of the gas mass $M_{gas}$ and the black hole mass $M_{bh}$. 

 For each model galaxy in our samples, we extracted its formation history which is the track along the main progenitor at each time step. In some cases, a galaxy will switch its identity from a central to a satellite or vice versa. This is due to the shortcoming of the SUBFIND algorithm \citep[][]{2017MNRAS.472.3659P}. So we eliminated the galaxies(at z=0) which are not always being central galaxies during their lifetimes from our original samples. Finally, about $38.6\%$ and $40.7\%$ members in original SampleF and SampleQ were removed, respectively.

\begin{figure}[ht!]
    \plotone{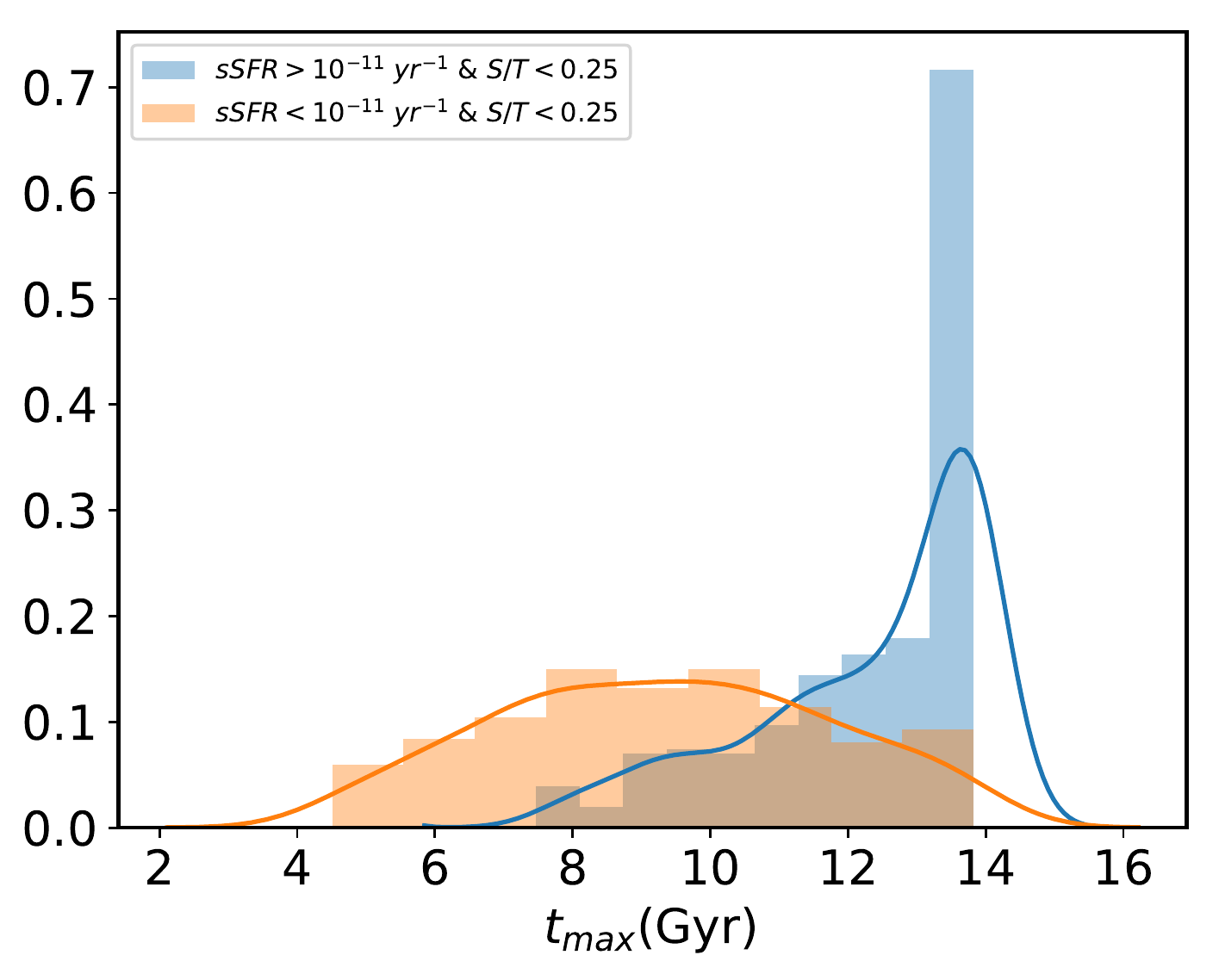}
    \caption{the distribution of $t_{max}$ for SampleQ(orange) and SampleF(blue) respectively.
    $t_{max}$ is defined as the time at which the galaxy have the largest gas mass $M_{gas}$ in the halo among the whole history. The blue distribution has a sharp narrow peak at the present time($z=0$), while the orange is very flat.}
    \label{fig:t_max}
\end{figure}

\begin{figure}[ht!]
    \plotone{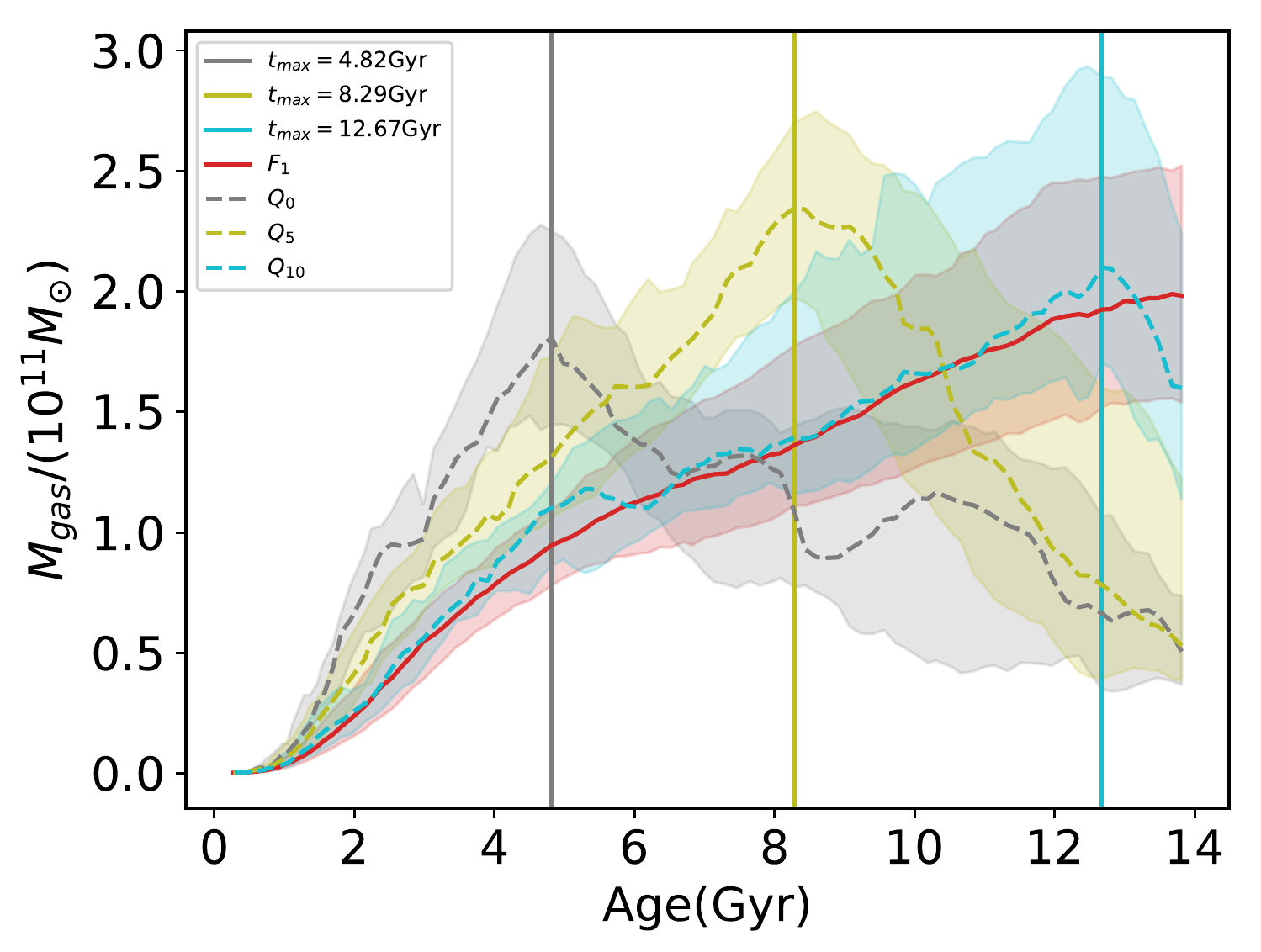}
    \caption{The evolution history of gas mass. The three quenching groups, $Q_{0}$, 
            $Q_{5}$ and $Q_{10}$, which correspond to those galaxies which achieved their peak gas mass in the Universe age interval $[4.29,5.12)\ Gyr$, $[8.28,9.06)\ Gyr$ and $[12.34,13.13)\ Gyr$, 
            correspond to the gray, green and light blue dashed lines respectively, while the red solid line represents 
            the star forming group, $F_{1}$($t_{max} \in [10.65,13.80)\ Gyr$). The three vertical lines shows the median of $t_{max}$
            for the corresponding quenching groups(with the same color). These values are $t_{max}\!=\!4.82 Gyr$ for $Q_{0}$, $t_{max}\!=\!8.29 Gyr$ for $Q_{5}$ and $t_{max}\!=\!12.67 Gyr$ for $Q_{10}$. The 25th percentile to
            75th percentile are denoted by the appropriately coloured corridors. It is notable that the gas mass of the quenched galaxies increase at first and decrease later, but for the star forming one, gas mass always rise. }
    \label{fig:gas_Mass}
\end{figure}

We then obtain the halo gas  mass along the main branch for each galaxy at $z=0$ in our sample. We found that for almost all galaxies in SampleQ, their halo gas  mass reached the maximum in the past at $z>0$ and then decreases up to $z=0$, which will be soon shown and explained more explicitly in Figure \ref{fig:gas_Mass}({ section \ref{sec:gas mass}}). So, for each galaxy in SampleQ and SampleF, we can define a new quantity, $t_{max}$, as the time when the halo gas  mass $M_{gas}$ reaches its maximum value during the formation history. Different galaxies have different $t_{max}$ and their distribution of the $t_{max}$ are plotted in Figure \ref{fig:t_max}. 

Obviously, the $t_{max}$ distribution of SampleF (blue histogram) has a sharp narrow peak at the present time($z=0$), while that of { SampleQ}  (orange histogram) is very flat. It means, for most SampleF galaxies, the gas mass keep increasing until $z=0$, however, for most galaxies of SampleQ, the gas mass peak appear before $z=0$. 


Considering the wide distribution of $t_{max}$ for SampleQ, we separated the samples into  sub-groups according their values of $t_{max}$. We divided the $t_{max}$ from $t=4.29\ Gyr$ to $t=13.8\ Gyr$ into 12 intervals: ([4.29,5.12), [5.12,5.88), [5.88,6.69), [6.69,7.45), [7.45,8.28), [8.28,9.06), [9.06,9.84), [9.84,10.65), [10.65,11.51), [11.51,12.34), [12.34,13.13), [13.13,13.80]{\footnote{these unevenly spaced time intervals are correspond to the evenly spaced snapshot intervals: [40,45), [45,50), [50,55), [55,60), [60,65), [65,70), [70,75), [75,80), [80,85), [85,90), [90,95), [95,99]}}$\ Gyr$). Then we got 12 sub-samples from SampleQ: $Q_{i,i=0,1\cdots 11}$ corresponding to those time intervals (which means the sub-group $Q_{0}$ consisted of galaxies with $t_{max} \in [4.29,5.12)\ Gyr$). The number galaxies of each sub-sample is 12, 14, 22, 28, 38, 31, 38, 36, 36, 24, 22, 22, respectively. 
Similarly, we separated SampleF into 2 groups: $F_{0}$($t_{max} \in [7.45,10.65)${ \footnote{correspond to snapshot interval [60,80)}}$\ Gyr$) of 70 members and $F_{1}$($t_{max} \in [10.65,13.80]${ \footnote{correspond to snapshot interval [80,99]}}$\ Gyr$) of 334 members since it has a sharp $t_{max}$ distribution.

In the following, we selected three representative sub-sample of SampleQ: $Q_{0}$($t_{max} \in [4.29,5.12)\ Gyr$),$Q_{5}$($t_{max} \in [8.28,9.06)\ Gyr$), $Q_{10}$($t_{max} \in [12.34,13.13)\ Gyr$) and one sub-sample of SampleF: $F_{1}$($t_{max} \in [10.65,13.80)\ Gyr$) to investigate their evolution history, as any two sub-samples with close $t_{max}$ values will have similar behaviour of their evolution.


\subsection{The evolution of the gas mass}\label{sec:gas mass}

We first show the evolution history of the gas mass $M_{gas}$. In Figure \ref{fig:gas_Mass}, the gray, green and light blue dashed line represent the three quenching sub-samples: $Q_{0}$, $Q_{5}$ and $Q_{10}$, respectively, while the red solid line represents the star forming one: $F_{1}$. The three vertical lines indicate the median value of $t_{max}$ (which are $t_{max}\!=\!4.82 Gyr$ for $Q_{0}$, $t_{max}\!=\!8.29 Gyr$ for $Q_{5}$ and $t_{max}\!=\!12.67 Gyr$ for $Q_{10}$.) of the corresponding quenching sub-samples(shown by the same color). The contour range represents the 25th percentile to 75th percentile distribution. It is seen that the halo gas  mass of all quenched samples reached their maximum at early stage then decrease with redshift. For the star forming galaxies, their halo gas  mass reach the maximum at the present day. This plot also shows that the halo gas  mass in SampleQ are lower than that in SampleF, consistent with what we have shown in Figure \ref{fig:halo}.  

\begin{figure}[ht!]
    \plotone{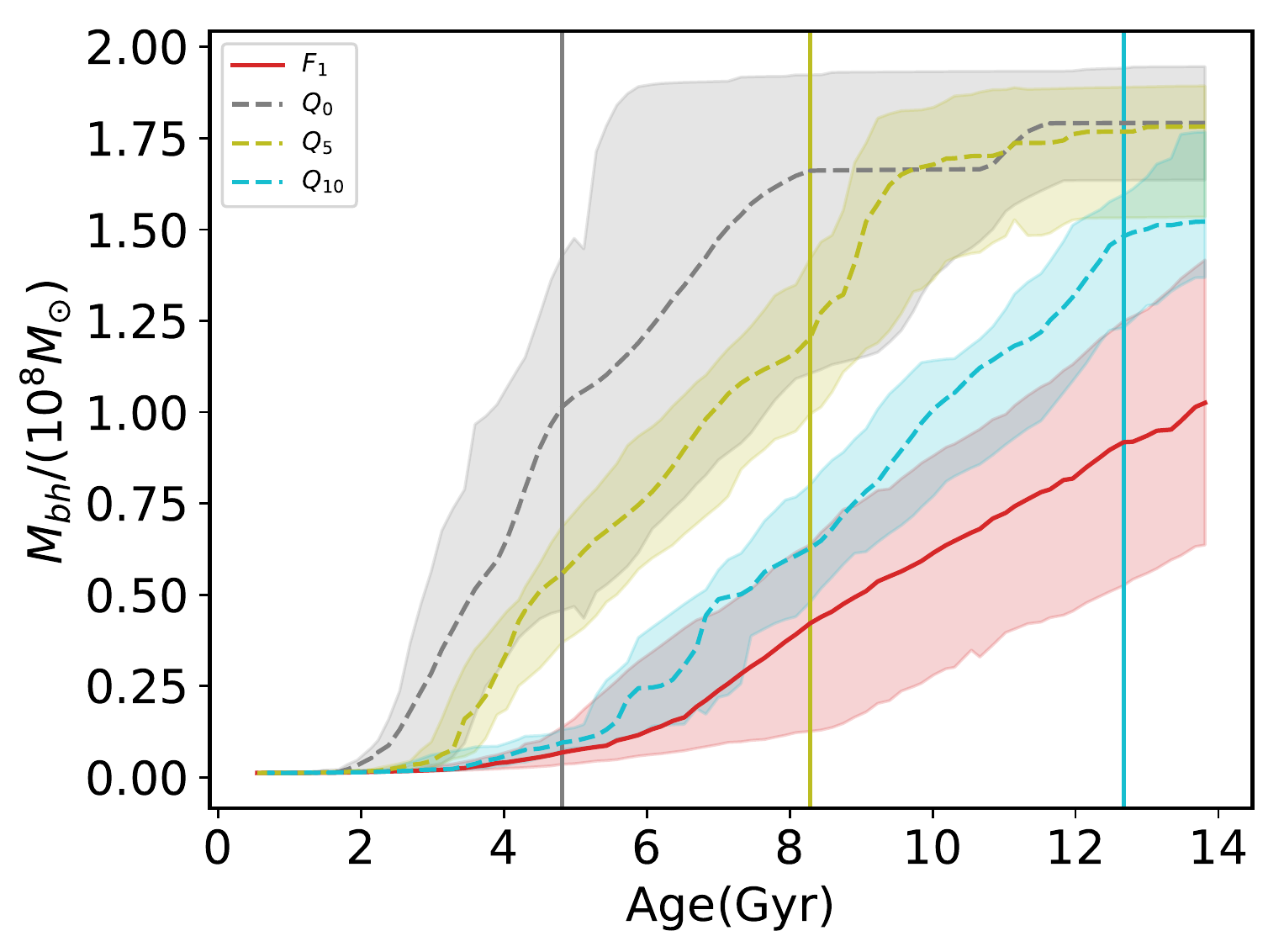}
    \caption{The evolution history of black hole mass. The meaning of the other signs are same as Figure \ref{fig:gas_Mass}. The $M_{bh}$ of the quenched samples grow faster and are larger than the star forming one.}
    \label{fig:bh_Mass}
\end{figure}

\subsection{The evolution of black hole mass}\label{sec:blackhole_mass}

We now show the evolution history of the black hole mass, $M_{bh}$, in Figure \ref{fig:bh_Mass}. All the notations are same as in Figure \ref{fig:gas_Mass}. The first signature to see is that three dashed lines are always above the solid line, which means the mass of $M_{bh}$ of the quenched samples grow faster and are larger than the star forming ones. Among the quenching samples, $M_{bh}$ of sub-sample with lower $t_{max}$ (e.g. $Q_{0}$) increase more rapidly and reached the threshold (
$\sim 10^{8}M_{\odot}$
, above which the feedback from black hole will switch from thermal to kinetic mode) earlier than that of the sub-sample with larger $t_{max}$ (e.g. $Q_{10}$). We can also find that the black hole mass $M_{bh}$ of all three quenched sub-samples reached the threshold almost at their own median $t_{max}$ (i.e. the time when the gas mass begin to decrease, marked by the vertical coloured solid line). The coincidence of time when black hole is above the threshold and when the total gas began to decrease clearly indicates it is the kinetic AGN feedback mode to transform a galaxy from star forming to quenching.

\begin{figure}[ht!]
    \plotone{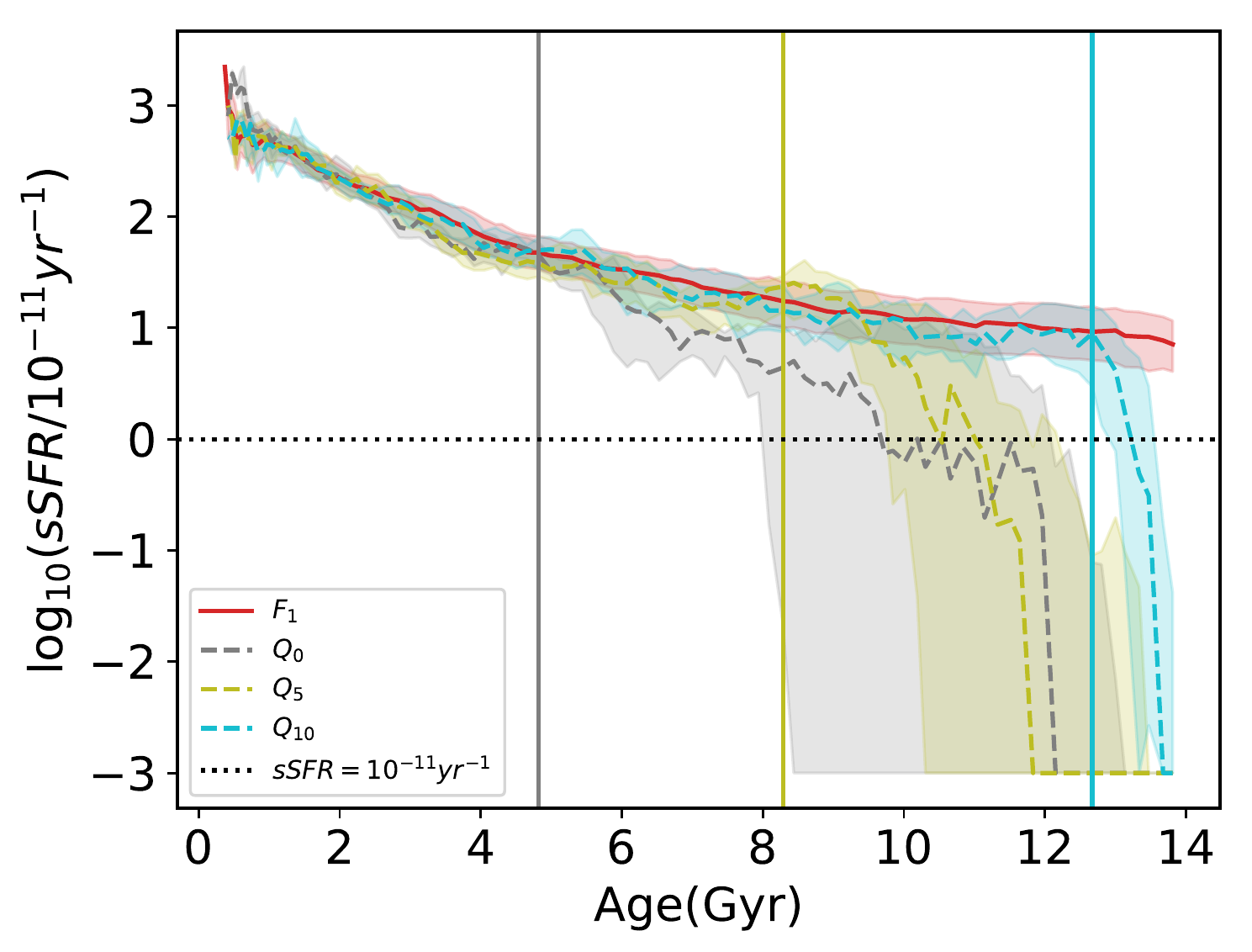}
    \caption{The evolution history of $sSFR$. The meaning of the other notations are same as Figure \ref{fig:gas_Mass}. It is notable that the dashed line (quenching groups) decrease more rapidly right after the relevant $t_{max}$. We set $\mathrm{sSFR} = 10^{-14}yr^{-1}$ for galaxies with unresolved SFR(i.e. $\mathrm{SFR}=0\  M_{\odot}yr^{-1}$)(which is different from the convention used in the Figure \ref{fig:ssfr_s/t}, just for convenience). The black dotted line is the criterion used to distinguish quenched($\mathrm{sSFR} < 10^{-11}yr^{-1}$) and star-forming($\mathrm{sSFR} > 10^{-11}yr^{-1}$) galaxies at $z=0$.}
    \label{fig:sSFR}
\end{figure}

\begin{figure*}[hbt!]
    \plotone{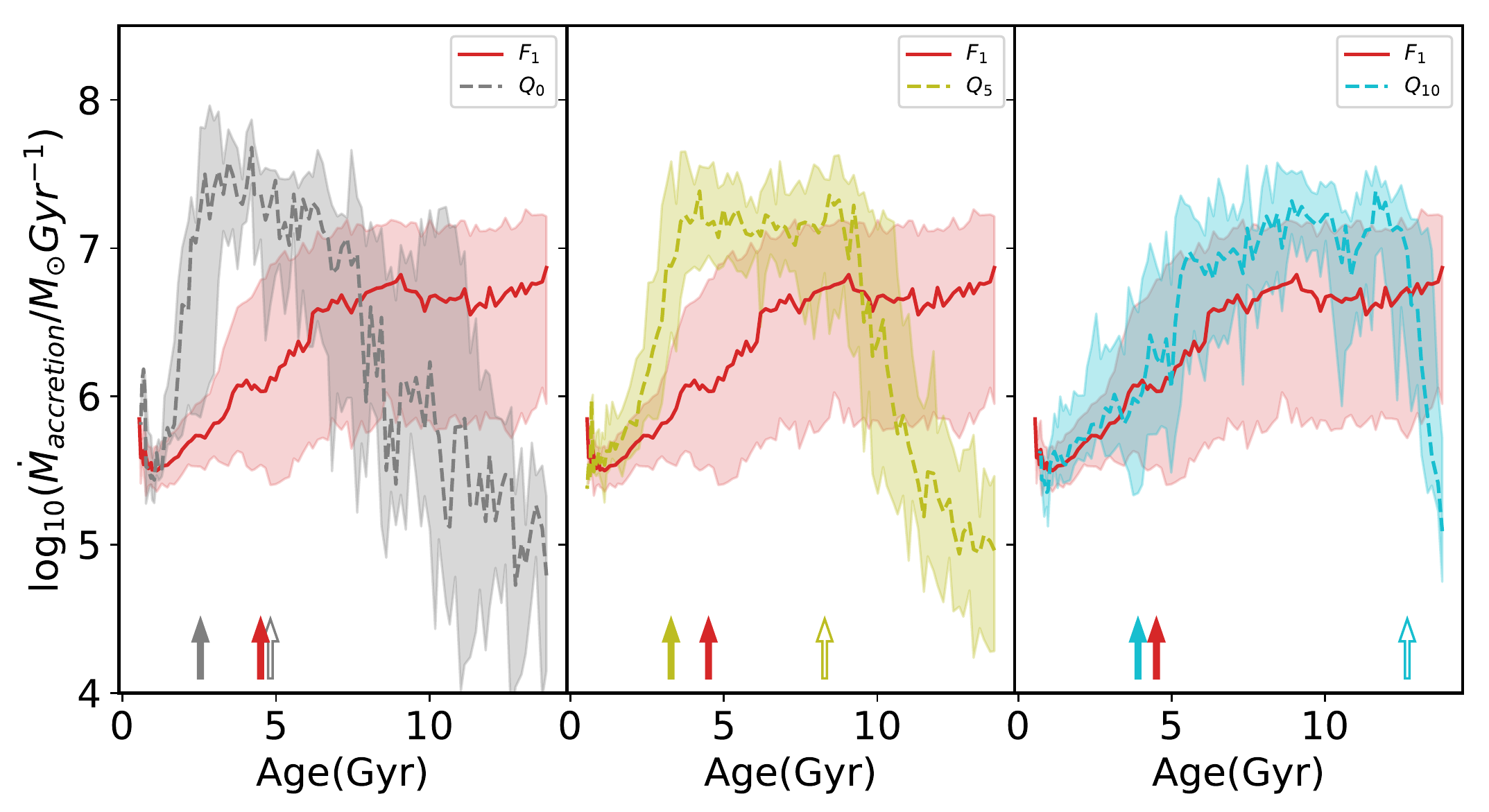}
    \caption{The evolution of black hole accretion rate. From left to right, the black hole mass accretion rate of each quenching sample group have been compared with that of the star forming one in each panel. We also shown the $t_{max}$ marked with open arrows.  The filled arrows have the same meaning with that in Figure \ref{fig:halo_mass}. 
    Quenched sub-samples keep very high accretion rates for a period of time then the rates decline rapidly. However for the star forming sample (red line), they keep relatively lower rates which are more or less constant.
    }
    \label{fig:M_accretion}
\end{figure*}

\subsection{The evolution of specific star formation rates }\label{sec:ssfr}
To further answer whether the decrease of gas mass directly leads to star formation quenching, in Figure \ref{fig:sSFR} we show the evolution of sSFR. The horizontal dotted line is the division for quenching and star forming. It can be found  that the $\mathrm{sSFR}$ of the quenched samples (the dashed lines) decreased more rapidly right after the corresponding $t_{max}$, and they will be quenched in 3$\sim$4 Gyrs. While for star forming ones (red solid line), sSFR had a relatively slow decrease and keep a relatively larger value ($>10^{-10} yr^{-1}$) above the threshold of quenching. This result indicates that the quenching of galaxies in SampleQ is really related to the loss of gas mass. Although some other process, such as stellar feedback or star formation could play an important role in consuming the gas, the comparison of the $Q_{10}$ (dashed blue line) and $F_{1}$ (red solid line) pointed out that the AGN feedback should play a key role in gas depletion, as they had almost same gas mass (see Figure \ref{fig:gas_Mass}) and sSFR evolution (Figure \ref{fig:sSFR}) before $t_{max}$, but with quite different black hole mass evolution (Figure \ref{fig:bh_Mass}). Only when galaxies have central black hole mass above the threshold of $M_{bh} \sim 10^{8}M_{\odot}$, the quenching from AGN feedback began to be effective. This conclusion also explains why some galaxies in SampleF with higher black hole mass ($\sim 10^{8}M_{\odot}$) are still star forming, as their black hole mass just reach this threshold very recently.

So we may conclude that the massive black hole and its effective kinetic feedback, could drive a large amount of gas loss thereby depleting the gas reservoir and thus effectively quenching star formation. In fact, the strong AGN feedback can also suppress gas inflow to halos which could also decrease the gas mass in halo\citep[e.g.][]{2020MNRAS.498.1668W}. { The overall effect of AGN feedback, either by producing outflow or suppressing inflow, is to decrease the total gas content of the galaxy, and in this work we do not distinguish these two processes.}

\section{How did black hole grow in massive disk galaxies?}\label{sec:grow-bh}
 We have found that the kinetic feedback of the SMBH is the key to quench star formation for massive disk galaxies. { It now comes to the question: how did the SMBHs grow in these massive disk galaxies? In the classical scenario of bulge-black hole co-evolution, the black hole and bulge growth is mainly through galaxies merger. However, this scenario dose not seem to apply to the current case: disk-dominate galaxies with massive SMBH. As mentioned in the introduction, this kind of galaxies do not follow the classical bulge mass-BH mass relation and their SMBHs mainly grows via non-major-merger processes. In the following section, we investigate how the SMBHs in our samples grow in the simulation.} 

\subsection{The possible path of black hole growth} \label{sec:black hole growth phase}

From the black hole growth evolution (Figure \ref{fig:bh_Mass}), we have found that the quenched sub-sample have experienced fast growth at early stage. We further investigate the evolution of the black hole mass accretion rate. Here the accretion rate is instantaneous black hole mass growth rate given by the TNG simulation. 
As shown in the Figure \ref{fig:M_accretion}, { during the interval between the open arrow and the solid one with the same color code}, the SMBHs of all 3 quenched sub-samples kept  very high accretion rates ($> 10^{7} M_{\odot}Gyr^{-1}$) for a period of time then accretion rates decreased rapidly. While for the star forming sample (red line), they kept relatively lower accretion rates ($\sim 10^{6.5} M_\odot Gyr^{-1}$).
{ The fact that both the quenched and star-forming samples experience such high accretion phase for several Gyrs may suggest that SMBH accretion play a dominant role in driving the growth of SMBHs in the massive disk galaxies. By using TNG300 simulation,  \cite{2018MNRAS.479.4056W} have obtained the same results as in this work, in which they found that SMBH($< 10^{8.5}M_{\odot}$) mainly grows via accretion.}

We also found that for quenched sub-sample the time when the accretion rate of black hole began to decrease is correlated with $t_{max}$ (marked with open arrows).  At around $t_{max}$ when the black hole mass reached the threshold($\sim 10^{8}M_{\odot}$), AGN feedback mode is switched from thermal to kinetic, so the surrounding gas is quickly expelled and it leads to lower accretion rate afterwards. 

This black hole growth pattern supports results from some previous work { \citep[][from EAGLE simulation]{2017MNRAS.468.3395M,2018MNRAS.481.3118M}}  which indicated that there are three different black hole growth phase: the stellar feedback regulated phase, the non-linear black hole growth phase and the AGN feedback regulated phase.
\cite{2017MNRAS.465...32B} found 
that the switch between the two phases are governed by the halo virial mass. When the halo virial mass is lower than certain critical mass $M_{crit}$, the 
mass accretion could be suppressed by the stellar feedback. When the halo virial mass is above the $M_{crit}$, the black hole would be in non-linear growth phase. The $M_{crit}$ can be calculated as  $M_{crit}=10^{12}(\Omega_{m}(1+z)^{3}+\Omega_{\Lambda})^{-1/8}M_{\odot}$. 

\begin{figure}[ht!]
    \plotone{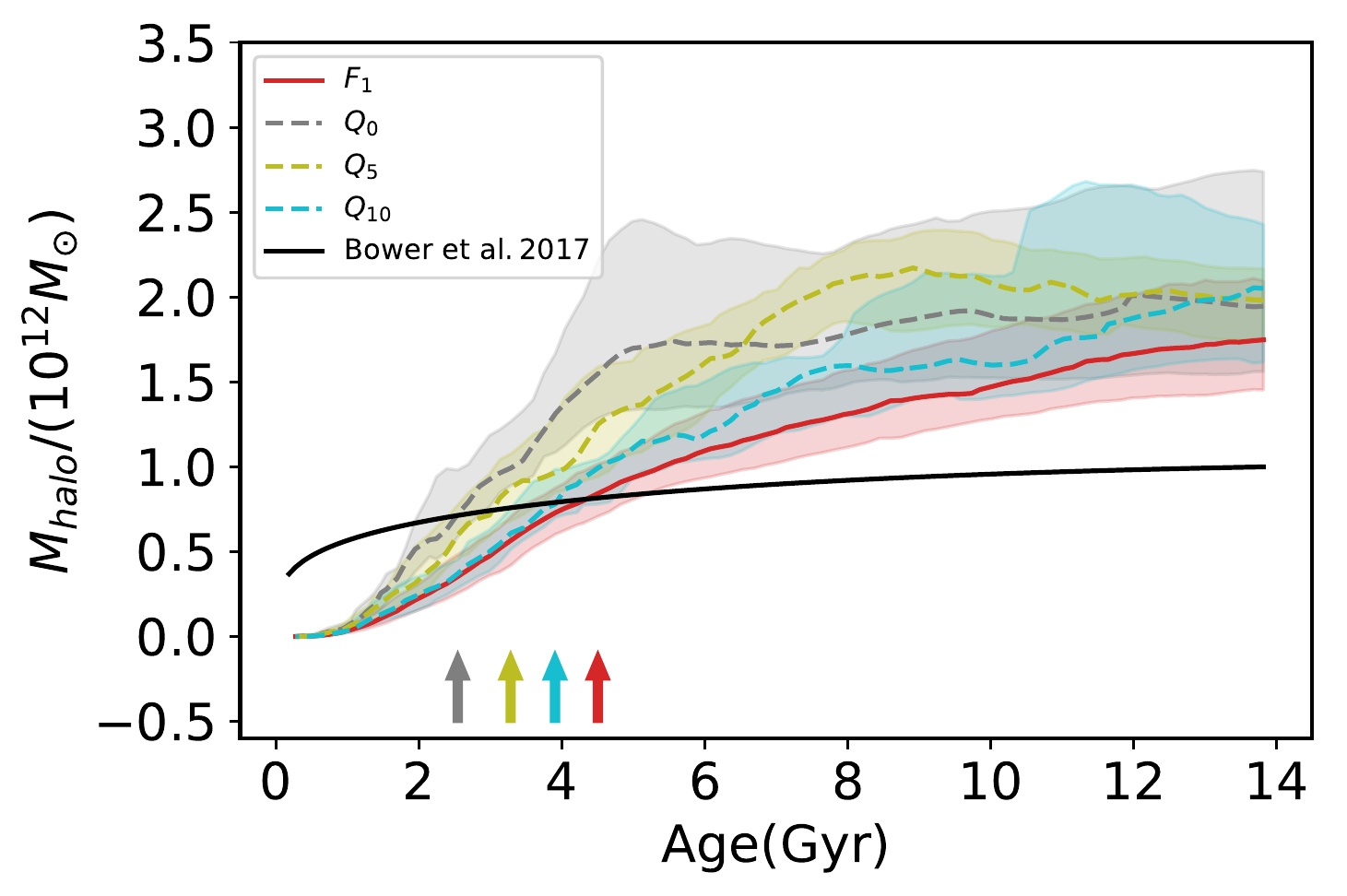}
    \caption{The evolution history of halo virial mass. The meaning of the other notations are same as Figure \ref{fig:gas_Mass}. The black line denotes the evolution of the critical halo virial mass $M_{crit}$, below which the black hole accretion will be suppressed by the stellar feedback \citep{2017MNRAS.465...32B}. the 
    coloured arrows represents the earliest snapshot at which the mass of the halo of the corresponding samples are higher than the the critical halo virial mass $M_{crit}$.}
    \label{fig:halo_mass}
\end{figure}

We show the halo virial mass evolution history of all 4 sub-samples, and  $M_{crit}$ (black solid line) in Figure \ref{fig:halo_mass}. We marked the critical time $t_{crit}$ when the median halo virial mass of each sub-sample exceeds the $M_{crit}$ with colored arrows. For those four sub-sample, the $t_{crit}$ are: $t_{crit}=2.54\ Gyr$\ (for $Q_{0}$), $t_{crit}=3.28\ Gyr$\ (for $Q_{5}$), $t_{crit}=3.9\ Gyr$\ (for $Q_{10}$), $t_{crit}=4.5\ Gyr$\ (for $F_{1}$). 

We also marked these critical time on the evolution of black hole accretion rate using solid arrows in Figure \ref{fig:M_accretion}. We found that for all the quenched sub-samples, the black hole has a high accretion rate during the time between $t_{crit}$ and $t_{max}$ (solid and open arrows in { Figure \ref{fig:M_accretion}}, respectively) well corresponding to the non-linear growth phase. 
After $t_{max}$, { black hole growth} is in AGN feedback regulated phase. During this phase, although the accretion rate is lower (in fact it still could be $> 10^5 M_{\odot}Gyr^{-1}$), the AGN feedback is still strong enough to decrease the star formation since kinetic mode is more effective than thermal mode to remove the surrounding gas.  
We note that at $z=0$, all of the quenched samples are in the AGN feedback regulated phase, while the star forming sample is still in the non-linear black hole growth phase as their black hole mass just approached the threshold mass ($\sim 10^{8}M_{\odot}$). 

\begin{figure}[ht!]
    \plotone{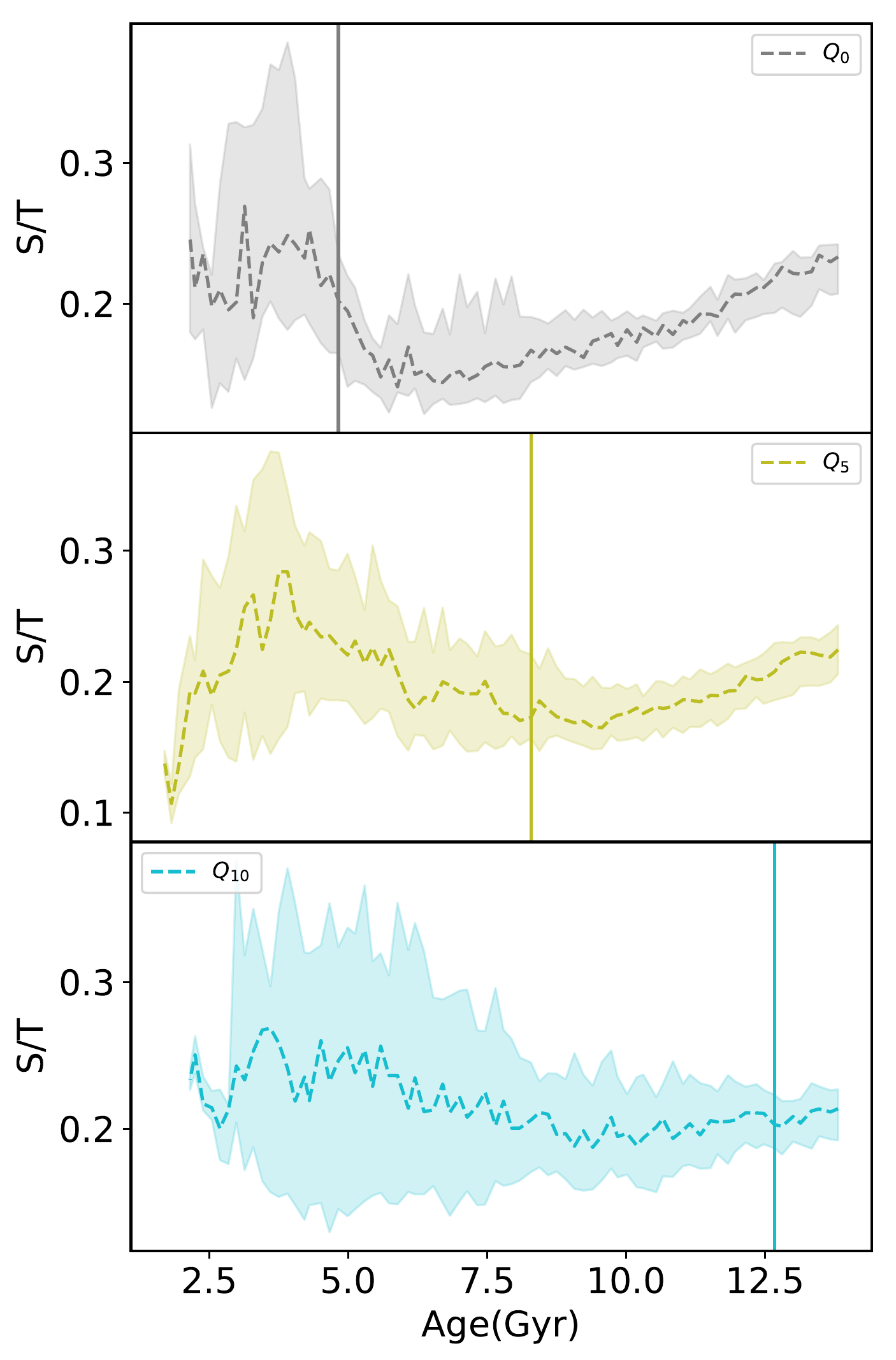}
    \caption{The evolution of the $S/T$ of the quenched groups. From top to bottom, we consider
    different quenched groups. We only choose the galaxies with at least 1000 star particles to construct the curve at every snapshot. Where the vertical lines is same as the one plotted on the Figure \ref{fig:gas_Mass}. These curves indicate that quenched massive disk galaxies got disk morphology many Gyrs ago.}
    \label{fig:ratioS_T}
\end{figure}

\begin{figure*}[p]
    \centering
    \includegraphics[scale=0.7]{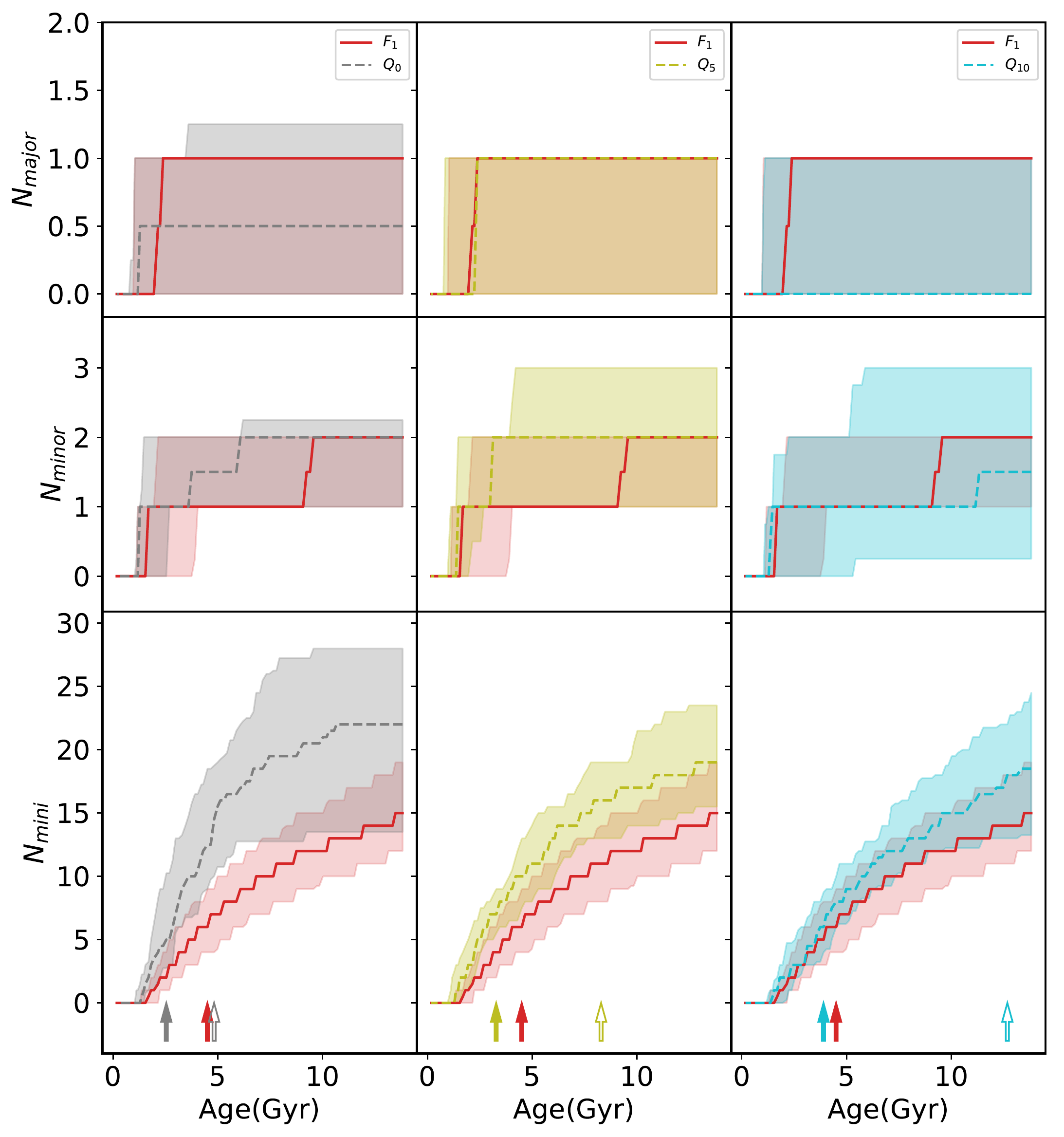}
    \caption{ We compare the cumulative merger numbers of each quenched samples and star forming ones on each column. From top to bottom, we consider three kinds of mergers:
    major merger($\mu \geq \frac{1}{3}$), minor merger($\frac{1}{3}>\mu \geq \frac{1}{10}$)
    and mini merger($\frac{1}{10}>\mu$). The meaning of the colored arrows is same as Figure \ref{fig:M_accretion}. All samples have almost one major merger occurred at very early time. While quenched samples experienced  more frequent mini mergers. As for minor merger, there are no distinction between star forming group and quenched ones.}
    \label{fig:merger}
\end{figure*}

\subsection{The evolution history of morphology and merger}
One key question about quenched disk galaxies is when they are formed. The recent study by \cite{2020arXiv201113749Z} using the integral field spectroscopy from the SDSS-IV MaNGA sample showed that massive red disk galaxies formed as early as elliptical galaxies, with more than half stellar mass are formed at 10 Gyrs ago. On the other hand, it is often thought that massive black hole is formed mainly in major merger, { under the merger-driven scenario for the co-evolution of the black hole and galaxy,} but major mergers usually lead to the formation of massive galactic bulge. Why do these massive disk galaxies lack significant bulge if their massive black hole are formed in major mergers? In the following we investigate the morphology evolution of these quenched disk galaxies and the role of merger for the formation of the disks. \par

Along the evolution track of each model galaxy, we decompose their star distribution into bulge and disk using the kinematic method mention in Section \ref{sec:sample}. We then plot the evolution of $S/T$ of quenched sub-samples in Figure \ref{fig:ratioS_T}. Here, considering the star mass resolution, we only selected galaxies with more than 1000 particles within $3\mathrm{R_{e}}$ to derive reliable estimate of $S/T$ at every snapshot. It is interesting to see that the median $S/T$ has a low value ($<0.25$) during the evolution, showing that those quenched massive disk galaxies have disk morphology for a long time.
 At early times the scatter is slightly larger, possibly due to the resolution effect (less star particles at early times), but after $t_{max}$ the scatter of $S/T$ is lower. \cite{2021arXiv210509722Z} also found that some massive disk galaxies preserve the disk morphology since their formation in TNG simulation. They argued that major mergers with special orbits could produce disk-dominant remnants. 

Usually, the formation of bulge is mainly due to major merger of galaxies. This long-lasting disk morphology of SampleQ may hint that major merger is not dominant during their formation. So we then study the merger event of our four sub-samples. Here, we classify three kinds of mergers according to the merger mass ratio: major merger($\mu \geq \frac{1}{3}$), minor merger($\frac{1}{3}>\mu \geq \frac{1}{10}$) and mini merger($\frac{1}{10}>\mu$). The mass ratio, $\mu$, defined as $\mu\equiv \frac{M_{\star,1}}{M_{\star,2}}$, where 
$M_{\star,1} < M_{\star,2}$ represents stellar mass of two merging galaxies respectively.(The definition of the mass ratio $\mu$ is same as \cite{2015MNRAS.449...49R}). Then we calculated the cumulative merger numbers for every galaxy.\par

In Figure \ref{fig:merger}, we compared the cumulative merger numbers of each sub-samples for 3 different merger types separately. As we can see from the first row, all the samples have at most one major mergers at early stage $t<t_{max}$. Interestingly, these major mergers did not quench the star formation of those galaxies (sSFR decreased significantly only after $t_{max}$, see Figure \ref{fig:sSFR}). This is consistent with the results of  \cite{2018MNRAS.479.4056W,2021arXiv211207679P} that major merger event is not a main cause of quenching in the TNG simulation.

From the second row of the Figure \ref{fig:merger}, it seems that there are no clear distinction between the number of minor mergers between the quenched samples and the star forming sample. However, for the mini merger shown in the lowest row, the quenched samples had more frequent mini merger events. We also marked $t_{crit}$ and $t_{max}$ by the solid and open arrows in the Figure \ref{fig:merger}. Most of the mini mergers happened between $t_{crit}$ and $t_{max}$. This is extract the period when the black hole has higher accretion rate, i.e, the non-linear black hole growth phase shown in the Figure \ref{fig:M_accretion}.
Furthermore, these frequent mini mergers nearly did not change the morphology of those galaxies. {Hence, we could conclude that the frequent mini mergers cause the rapid growth of SMBHs in quenched disk galaxies and these mini-mergers did not significantly change the disk morphology of the galaxies. All massive disk galaxies in our sample, either quenched or star-forming, have experienced few major mergers. Our conclusion  of non-major-merger driven  growth of SMBHs in disk galaxies is consistent with many recent studies \citep[e.g.][]{2012ApJ...744..148K,2014MNRAS.440.2944K,2018MNRAS.476.2801M,2020MNRAS.494.5713M}.}  \par

\section{summary and conclusion} \label{sec:conclusion}
The recent finding of quenched massive disk galaxies has raised great concern about how they formed and were being quenched \citep[e.g.][]{2019ApJ...884L..52Z,2021ApJ...911...57Z,2020arXiv201113749Z,2021SCPMA..6479811X}. Most current work focus on their observational properties and little attention is paid to their origins. In this work, we used the TNG300-1 simulation to study the formation and quenching of massive ($M_{\star}>10^{10.5}M_{\odot}$) disk galaxies in detail. 

We selected massive ($M_{\star}>10^{10.5}M_{\odot}$) central galaxies at $z=0$ from TNG. By comparing quenched sample ($\mathrm{sSFR}< 10^{-11} yr^{-1}\mathrm{and}\ \mathrm{S/T}<0.25$) and star forming sample($\mathrm{sSFR} > 10^{-11} yr^{-1}\mathrm{and}\ \mathrm{S/T}<0.25$), we found that quenched sample had less total gas and lower hydrogen density, but they posses more massive black hole than star forming sample. 

Then we investigated the evolution history of those massive disk galaxies. We found that for those quenched galaxies, their halo gas  mass and $\mathrm{sSFR}$ began to decrease when the mass of SMBH reached the threshold ($\sim 10^{8} M_{\odot}$), where the feedback mode of the SMBH will be changed from thermal to kinetic mode which is more efficient to expel gas from the galaxy.

We further investigated the accretion history of black hole and found that the black hole accretion rate of quenched samples is higher than star forming counterparts, during the period between $t_{crit}$ and $t_{max}$. This period of high accretion rate corresponded to the non-linear black hole growth phase. From the evolution of 
morphology and merger event,  { we found those quenched galaxies have formed their disk when $\mathrm{Age(cosmic)}\lesssim t_{max}$ and}  being able to keep the disk morphology for a long time. Both quenched and star forming samples lack major mergers. Compared to the star forming galaxies, the frequent mini merger ($\mu <\frac{1}{10}$) in quenched galaxies during the time interval ($t_{crit}<\mathrm{Age(cosmic)}<t_{max}$) lead to  rapid growth of black hole while preserving their disk morphology.  

Therefore we can conclude that in the IllustrisTNG simulation, the frequent mini mergers could make the SMBHs grow rapidly, 
and the kinetic feedback of black hole produced low cold gas density and finally quenched the massive disk central galaxies.\par


Under this quenching scenario, we should expect that the massive quenched disk central galaxies
will hold a massive black hole and own a low gas fraction($f_{CGM}\equiv \frac{M_{gas}}{M_{200}}$). However the gas fraction $f_{CGM}$ is not a direct observable. {In order to test our result, we need to find a direct observable accessible with current technologies and telescopes to estimate $f_{CGM}$. Recently, By using EAGLE simulations,  \cite{2020MNRAS.491.2939O} argued that, for $L^{\star}$-mass $(M_{\star}=10^{10.2}-10^{10.7}M_{\odot})$ central galaxies, the covering fraction, $C_{\mathrm{C_{IV}>13.5,100kpc}}$, of absorbers $\mathrm{C_{IV}}$ with column density $N_{\mathrm{C_{IV}}} > 10^{13.5} cm^{-2}$ within an impact parameter $100 kpc$ is a well observational proxy for $f_{CGM}$. They found that $C_{\mathrm{C_{IV}>13.5,100kpc}}$ correlates strongly and positively with $f_{CGM}$(the figure 3 of their paper). More importantly, The $\mathrm{C_{IV}}$ 1548, 1551 $\mathring{A}$ doublet is obtainable via the Cosmic Origins Spectrograph on the Hubble Space Telescope in the local universe. Thus, for local galaxies, the $C_{\mathrm{C_{IV}>13.5,100kpc}}$ are available and their SMBHs mass are also accessible. we hope that massive and quenched disk galaxies owning low covering factor $C_{\mathrm{C_{IV}>13.5,100kpc}}$ and giant SMBH could be used to test our conclusions. }

\acknowledgments
We thank Yaoxin Chen for helpful comments. This work is supported by the  NSFC (No. 11861131006,11825303, 11703091,11333008) and the 973 program (No. 2015CB857003). We acknowledge the science research grants from the China Manned Space project with NO. CMS-CSST-2021-A03, CMS-CSST-2021-A04.

\bibliography{QMSG}{}
\bibliographystyle{aasjournal}

\end{CJK*}
\end{document}